\documentclass[superscriptaddress,twocolumn,floatfix,nofootinbib]{revtex4-2}
\usepackage{times,fancyhdr}
\usepackage[dvips]{graphicx}
\usepackage{amsmath,amssymb,bm,dsfont}
\usepackage{mathrsfs}
\usepackage{mathtools}
\usepackage{setspace}
\usepackage{color,soul}
\usepackage{hyperref}
\usepackage{tikz}
\usetikzlibrary{arrows.meta,positioning,calc,decorations.pathmorphing}

\def\beq{\begin{equation}}
\def\eeq{\end{equation}}
\def\beqn{\begin{eqnarray}}
\def\eeqn{\end{eqnarray}}

\begin{document}
\title{How Gravity Can Explain the Collapse of the Wavefunction}
 
\author{Sabine Hossenfelder}
\affiliation{is looking for a new affiliation.}

\begin{abstract}
I present a simple argument for why a fundamental theory that unifies matter and gravity gives rise to what seems to be a collapse of the wavefunction. The resulting model is local, parameter-free and makes testable predictions.  
\end{abstract}

\maketitle

\section{Introduction}

The measurement problem in quantum physics concerns the question of why we generically do not observe the outcome of the Schr\"odinger evolution, but merely one eigenstate of the measurement observable with a probability that can be computed from the wavefunction. While we can mathematically describe this process with the reduction (or `update' or `collapse') of the wavefunction, the collapse is then not local, which is difficult to reconcile with general relativity. The question of why we never observe macroscopic superpositions (of measurement eigenstates) therefore turns into the question of how the process to arrive in a measurement eigenstate can happen locally. 

Several models have been proposed to convert the sudden collapse of the wave function into a gradual, physical process. The most prominent among them are the Ghiradi-Rimini-Weber ({\sc GRW}) model \cite{ghirardi1986unified} and related approaches \cite{tumulka2006relativistic}, Di\'{o}si's model of stochastic gravitational collapse \cite{diosi1987universal,diosi1989models},  as well as Penrose's \cite{penrose1996gravity,penrose1998quantum} model of gravitationally induced collapse. These models, however, are all non-local in the sense of violating Bell's condition of local causality. The purpose of this paper is to develop a local collapse model.   

We know from Bell's theorem \cite{Bell1964OnEPR,Bell2004Speakable} that any locally causal model that correctly describes observations needs to violate measurement independence. Such theories are sometimes called `superdeterministic' \cite{Hossenfelder2020Rethinking,Hossenfelder2020Perplexed}. It is therefore clear that to arrive at a local collapse model, we must use a superdeterministic approach. The aim of this paper is to show that a local collapse arises in a superdeterministic setting from what I believe to be fairly general assumptions about quantum gravity.  

In the following $\hbar = c = 1$. 

\section{Model Definition}
\label{sec:model}
\subsection{Geometry and Matter}
\label{ssec:geon}

As starting point, I want to use an old idea for the unification of the fundamental interactions: that matter, radiation, and geometry are ultimately the same. 

One way to interpret this statement is that the particles in the standard model are really geometrical in nature: stable, noise-free, subsystems. This idea was pursued for example by John Wheeler with the geon approach \cite{wheeler1955geons}, but it survives in more modern formulations such as Spinor Gravity \cite{hebecker2003spinor}, geometric engineering in string theory \cite{katz1997geometric}, or braided spin networks \cite{bilson2012emergent}. 

Another way to interpret the statement is that geometry is a purely relational property that arises entirely from matter. This, too, is not a new idea; it has been pursued for example in the context of Causal Fermion Systems \cite{finster2015causal}, Shape Dynamics \cite{barbour2012shape}, or Geometric Unity \cite{Weinstein2021GUsite}.

Classical gravity already has these features to some extent. On the most trivial level, we can infer the mass and charge of a particle from its gravitational field. Furthermore, there are known links between the solutions of Yang-Mills theory and those of gravity \cite{monteiro2014black}. A difficulty here is to account for quantum properties like fractional spin. Nevertheless, it is clear that gravity carries a lot of the information from the particle sector already.

What I am assuming here is then that in the to-be-found underlying theory, geometry carries the same information as the particles because they {\emph{are}} the same. Gravity is in this sense fundamentally different from the other interactions: The electromagnetic interaction, for example, does not carry any information about the mass of the particles. Yet gravity carries information about the particles' charges. We may note in passing that this would solve the black hole information loss problem. 

Concretely, I will take this idea to imply that we have a fundamental quantum theory in which particles and their geometry are one and the same quantum state. That is, the geometry is fully determined by the particles' properties, and vice versa. There are no extra degrees of freedom. I am here including gravitons as a type of particle, one for which the relation is particularly obvious. 

To be even more concrete, let us call this fundamental quantum state $|\Psi \rangle$. We then want to recover the familiar Hilbert space of quantum gravity ${\mathscr{H}}$, that is a product of matter and geometric degrees of freedom
\beqn
\mathscr{H} = \mathscr{H}_{\rm m} \otimes \mathscr{H}_{\rm g}~.
\eeqn
The geometric degrees of freedom do not necessarily have to be the metric. They could be other geometric properties, like the connection, loops, networks, or anything else that, in the fundamental theory, might replace space-time. 

The assumption that I have made, that matter and geometry are ultimately the same, means we will not get the full Hilbert space ${\mathscr{H}}$, but rather a subset of product states $\mathscr{M} \coloneq \{|\Psi\rangle \otimes U |\Psi \rangle \}$. I have here introduced a unitary operation $U$ that accounts for the possibly different identification of degrees of freedom. We correspondingly assume that the underlying theory has a total Hamiltonian $\hat H_{\rm tot}$ that acts on each factor separately and is of the form
\beqn
\hat H_{\rm {tot}} = \frac{1}{2} \left( \hat H_{\rm u} \otimes \openone + \openone \otimes U \hat H_{\rm u} U^{\dag}  \right)~, \label{eq:htot}
\eeqn
where $\hat H_{\rm u}$ is the unknown underlying Hamiltonian and $\openone$ is the identity operator.  

The reason for this reduced Hilbert-space, $\mathscr{M}$, is that an entangled state between matter and geometry would have extra information in the phases that are neither in the matter nor in the geometry sector. But by assumption each of those sectors already carries the full information. 

This seemingly innocent assumption causes an immediate problem, which is that in the best understood approaches to quantum gravity---perturbatively quantised and canonically quantised gravity---the Hamiltonian evolution generates entanglement between matter and geometry. We therefore need to reconcile these two approaches. 

One might at this point say, well, this discrepancy just serves to show that we cannot assume a product state! But as I will argue below, this might be the reason why we do not have a physical description of the measurement process in quantum mechanics.

\subsection{Off the Hamiltonian}
\label{ssec:Off}

We will reconcile the product assumption with the canonical quantum gravity evolution by conceding that the time-evolution of the product state cannot be a solution to the Schr\"odinger equation. We will further assume that this deviation from the Schr\"odinger evolution leads to a suppression of the transition amplitude. That is, we allow the evolution of the state in Hilbert space to proceed unitarily\footnote{This means in particular that states are normalised to 1 as usual.} on any path that evolves locally and by the known interaction terms, and that respects all conservation laws, but it needs to fulfil the product state constraint.

Before we move on, I want to add some words on how to interpret this approach. I have started from the assumption that we have an underlying fundamental theory that unifies matter and geometry. This underlying theory has its own dynamical law that, alas, I don't know. The product state constraint will not give us the underlying dynamical law. It is rather a phenomenological model to explain our observation of the collapse process that will, hopefully, allow us to identify ways to test the underlying theory.

This is similar to how, in the early days of quantum mechanics, physicists postulated discrete atomic energy levels and a fractional electron spin just because that agreed with observations, despite not knowing the underlying law. The underlying laws were then discovered, and the theory developed, from further studies of the observed phenomena. 

It is straightforward to estimate how large the deviation of a product path $|\Psi \rangle $ from the usual Schr\"odinger evolution $|\Psi' \rangle$ is. First we may note that since the product state concerns gravitational degrees of freedom, we expect the product constraint to only make a difference when quantum gravitational effects are taken into account. That is, even without doing any calculation, we expect the effect to be small for states with small masses/energies.

To calculate the deviation, we then define the residual $R \coloneq ( i \partial_t - \hat H) | \Psi \rangle $ with respect to the standard Hamiltonian $\hat H$ (i.e.\ the known one of, say, canonical quantum gravity, and not the unknown $\hat H_{\rm u}$) and use the functional
\beqn
S \coloneq \int dt ~ ||R|| ~\label{eq:res}
\eeqn
to integrate the deviation from the Schr\"odinger evolution, where, as usual, $||\cdot || = \sqrt { \langle \cdot | \cdot \rangle }$.

It will be helpful in the following to decompose the integrand into two components, one that is perpendicular to $|\Psi \rangle$, that we will call $|R_\perp\rangle$, and one that is parallel to it, that we will call $|R_\parallel \rangle$. We obtain them as usual as projections 
\beqn
|R_\perp\rangle &=& (\openone - |\Psi\rangle\langle \Psi| ) |R\rangle~, \\
|R_\parallel \rangle &=& \langle \Psi | R\rangle | \Psi \rangle ~.
\eeqn
Then we have
\beqn
||R||^2 = ||R_\perp||^2 + ||R_\parallel||^2~, \\
||R_\parallel||^2 = |\langle \Psi | R \rangle|^2~.
\eeqn

Since both terms are strictly positive, they must separately be minimal. The parallel component, however, can always be set to zero by multiplying $|\Psi \rangle$ by a suitable time-dependent phase. Suppose we have $\langle \widetilde \Psi | \widetilde R \rangle \neq 0$. Then instead choose $|\Psi\rangle = e^{-i \varphi(t)} |\widetilde \Psi \rangle$, where 
$\dot \varphi = - \langle \widetilde \Psi | \widetilde R \rangle$,
and then you have $\langle \Psi |  R \rangle = 0$. This is sometimes called the ``energy gauge.'' 

The word ``gauge'' is somewhat misleading in our case, since the residual actually does depend on this phase, so it is not a gauge. Still, I will use this term in the following. 
The energy gauge is a shortcut to finding the minimum of the functional. One could alternatively just use $||R_\perp||$ for the functional and assume the energy gauge.  

\subsection{Teleology}
\label{ssec:Teleology}

In my mind the biggest shortcoming of Penrose's argument for gravitationally induced collapse is that once the gravitational self-energy is large enough to make a non-negligible contribution to the time-evolution, it is too late to collapse the state locally. This problem is common to all decoherence-based approaches, and it is here where superdeterminism comes in. 

Superdeterminism is mathematically defined as a correlation between the measurement settings $X$ and the presumed to exist hidden variables $\lambda$, i.e.\ $\rho(X|\lambda) \neq \rho(X)$, where $\rho$ is the probability distribution of the hidden variables. However, as we have argued in \cite{Hance2021Ensemble}, what this really means is that the time-evolution of the underlying state depends on the measurement setting: They must fit together. Superdeterminism is therefore best understood as a constraint on possible evolutions (an all-at-once constraint in the terminology of \cite{adlam2024taxonomy}) or as a superselection rule on possible outcomes of the time-evolution.

One can assume that the outcome of the time evolution is always a detector eigenstate (as we did in \cite{donadi2020toy}) to get a model that reproduces quantum physics. But what one really wants is to explain what outcomes we get and why, and hence how the classical world emerges. The model presented here will go a long way towards this. 

As argued in \cite{donadi2021future}, the action principle is ideally suited to incorporate a superselection rule because technically it already depends on both initial and end states. In classical physics, the action principle is only seemingly teleological, by which I mean it only seemingly depends on a boundary condition in the future. This is because we can use the principle of least action to derive the Euler-Lagrange equations which merely require initial values, and no future input. 

However, the action principle opens a door to formalising a theory that violates measurement independence by not only using a variation over all possible paths, but also a variation over all possible end-states of the paths. Our first step at an action principle will thus simply be that the time-evolution which is realised is the one that is (a) a stationary state of the residual action Eq.\ (\ref{eq:res}), $\delta S=0$, and (b) has the minimal residual among all stationary states given our product ansatz. This will turn out to be not quite correct, but we will refine it later on. 

\subsection{Multipartite and Interacting Systems}
\label{ssec:mult}

It is clear that if this model is to describe the measurement process, then it must deal with macroscopically large systems that are composed of many particles. We therefore have to think about how to deal with systems that have multiple components.

Let first consider the case of two non-interacting systems in a product state $|\Psi\rangle = |A \rangle \otimes |B\rangle$ with Hamiltonian
\beqn
\hat H = \hat H_A \otimes \openone + \openone \otimes \hat H_B~.
\eeqn
The residual is 
\beqn
| R \rangle = |R_A \rangle \otimes |B\rangle + |A\rangle \otimes |R_B \rangle~,
\eeqn
where $|R_A \rangle = ( {\rm i} \partial_t - \hat H_A ) | A \rangle$ and $|R_B \rangle = ( {\rm i} \partial_t - \hat H_B ) |B \rangle$.

To fix the phases for the energy gauge, we set $\langle A | R_A \rangle = \langle B | R_B \rangle = 0$. Then 
\beqn
||R||^2 = ||R_A||^2 + ||R_B||^2~.
\eeqn
This expression generalises in the obvious way to any number of systems: $||R||^2$ will be a linear sum of the individual contributions. Consequently, the contribution of each can be minimised separately, and for $n$ identical separable subsystems, $||R||$ will scale with $\sqrt{n}$

If the systems are interacting, we add a term that couples them
\beqn
\hat H_{\rm int} \coloneq \sum_k \hat I^k_A \otimes \hat I^k_B~.
\eeqn
Then we define
\beqn
\langle V  \rangle &\coloneq& \langle A | \langle B | \hat H_{\rm int} |B \rangle |A \rangle ~, \\
\hat V_A &\coloneq&  \langle B | \hat H_{\rm int} |B \rangle  ~, \\
\hat V_B &\coloneq&  \langle A | \hat H_{\rm int} |A \rangle  ~, \\ 
|R'_A \rangle &\coloneq& \left( i \partial_t - \hat H_A - \hat V_A + \langle V \rangle~ \right) |A \rangle,\\
|R'_B \rangle &\coloneq& \left( i \partial_t - \hat H_B - \hat V_B + \langle V \rangle~ \right) |B \rangle, \\
\widetilde V &\coloneq& \hat H_{\rm int} - \hat V_A \otimes \openone - \openone \otimes \hat V_B + \langle V \rangle \openone~.
\eeqn
In the energy gauge, we then have
\beqn
||R||^2 = ||R'_A||^2 + ||R'_B||^2~ + \langle \Psi | \widetilde V^2 |\Psi \rangle ~.
\eeqn
That is, the functional for two interacting subsystems is the deviation of each subsystem from the mean field path, plus a contribution from the interaction.

\subsection{Generalisations}
\label{ssec:gen}

While I have here for simplicity used the Newtonian limit and first quantisation, the derivation could be extended to a generally covariant quantum field theory. This can be done as usual by replacing the Lagrangian by an integral over density operators
\beqn
{\cal{L}} = i \frac{\delta}{\delta \Sigma} - \hat{\cal H}_\nu N^\nu~,
\eeqn
where $N^\nu = (N, N^i)$ is the lapse/shift vector, $\Sigma$ is a family of Cauchy hypersurfaces, and $\hat{\cal H}_\nu$ is the constraint operator density \cite{dewitt1967quantum,teitelboim1973commutators,kieferbook}.  With that, the functional $S$ becomes a space-time integral that is by construction a scalar. 

That said, general covariance is naturally broken in a typical quantum experiment by the rest frame of the detector.

\section{Model Properties}

Having formulated the mathematical framework of the new model, I now want to explain what it is good for. 

\subsection{The Penrose Case}
\label{ssec:pen}

The case I want to look at first is one in which we generate a particle of mass $m$ and with wavefunction $|\chi\rangle $ in a superposition of two places (in the following called branches), $\vec x_1$ and $\vec x_2$ that we will call $|\chi_1 \rangle$ and $|\chi_2 \rangle$. That is, the wavefunction of the particle is $|\chi \rangle = \alpha_1 |\chi_1 \rangle + \alpha_2 |\chi_2\rangle$. 

Strictly speaking, the particle's wavefunction in each location has its own gravitational field. For simplicity, we will use the Newtonian limit, in which the branches each have their own Newtonian potential $\Phi_1$ and $\Phi_2$, or corresponding quantum states $|\Phi_1 \rangle$ and $|\Phi_2 \rangle$, respectively. We will also assume that the wave-packets do not overlap, so they are orthogonal, that each branch separately is to good approximation classical. This is the common setup used for example in \cite{penrose1996gravity}.  I will also assume that the branches are so far apart that $\Phi_1(\vec x_2) \approx \Phi_2(\vec x_1) \approx 0$. 

Please do not think of the particles as point-like as that will cause singular gravitational potentials. Better think of them as suitably smeared-out coherent states. 

The usual quantum gravitational evolution will generate an entangled state of the form
\beqn
|\Psi' \rangle = a'_1 (t) |\chi_1\rangle |\Phi_1 \rangle +  a'_2 (t) |\chi_2\rangle |\Phi_2 \rangle~,
\eeqn 
which we assume to satisfy the Schr\"odinger equation of canonical quantum gravity, whereas our product state schematically has the form
\beqn
|\Psi \rangle = \left( a_1 (t) |\chi_1\rangle  +  a_2 (t) |\chi_2\rangle \right) \left( a_1 (t) |\Phi_1 \rangle + a_2 (t)  |\Phi_2  \rangle \right)~,
\eeqn 
and fulfils the Schr\"odinger evolution in whatever is the underlying theory.

We can then estimate $S$ as follows. First we assume, as Penrose in his argument \cite{penrose1996gravity}, that we have gauge-fixed the metrics of the different branches. This means in practice that we can express them both as functions of the same coordinate system, and we can compare them pointwise. In general, this is a complicated procedure, but for two wave-packets that are each approximately classical, we will just be left with one Newtonian potential focused on the position of the first particle, and one focused on the other particle. 

Since $|\Psi' \rangle$ evolves under the usual Hamiltonian, each branch has a slightly different time evolution stemming from the slightly different lapse functions which stem from the different Newtonian potentials. This means that each branch picks up a different phase that we can write as
\beqn
|\Psi' \rangle = \alpha'_1 e^{ - {\rm i} t m \Phi_1} |\chi_1\rangle |\Phi_1 \rangle +  \alpha'_2 e^{ - {\rm i} t m \Phi_2} |\chi_2\rangle |\Phi_2 \rangle~,
\eeqn 
where we have discarded an overall phase from the mass as it's the same on both branches. In this expression, $\alpha'_1$ and $\alpha'_2$ are constant, and the time-dependence has entirely moved into the phases.

For our product state, we will use the same coefficients $\alpha_1 = \alpha'_1$ and $\alpha_2 = \alpha'_2$  to preserve the weights, but we note that the phases pick up a factor $1/2$ from the doubling procedure, see Eq.\ (\ref{eq:htot}). If we expand $|\Psi \rangle$, then the two unmixed factors $|\chi_1 \rangle | \Phi_1 \rangle$  and  $|\chi_2 \rangle | \Phi_2 \rangle$ will solve the canonical Schr\"odinger equation. The residual is therefore schematically of the form
\beqn
|R\rangle &=& m ~ \alpha_1 \alpha_2  ~ \exp{\left(- {\rm i} ~ t ~ m ~ \Phi_{12}/2 \right) }  \Phi_{12}   \nonumber \\
 &\times& \left(  |\chi_1 \rangle |\Phi_2 \rangle  +   |\chi_2 \rangle |\Phi_1 \rangle \right)/2 ~,
\eeqn
where $\Phi_{12} = \Phi_1 (\vec x_1) + \Phi_{2} (\vec x_2)$. From this we get
\beqn
\sqrt {|\langle R | R \rangle |} = \frac{1}{2} m |\alpha_1 \alpha_2| |\Phi_{12} | \sqrt{2}~.
\eeqn

However, this is not the minimal value of the functional $S$ because $|\Psi \rangle$ is not in the energy gauge. Integrating $\dot \varphi = \langle\Psi | R \rangle$, we get $\varphi = t m |\alpha_1 \alpha_2|^2 |\Phi_{12}|$, and so the minimal norm becomes
\beqn
\sqrt {|\langle R | R \rangle |} = \frac{1}{2} m |\alpha_1 \alpha_2| |\Phi_{12} | \sqrt{2 - 4 |\alpha_1 \alpha_2|^2}~. \label{eq:Pres}
\eeqn 

The value of the square root in Eq.\ (\ref{eq:Pres}) is $\in [1, \sqrt{2}]$ and just makes an order-one correction. Since $|\alpha_1| = \sqrt{1-|\alpha_2|^2}$ we see that, as expected, the residual vanishes if one has only one branch, because no entanglement is created. The only other case when the expression vanishes is when the gravitational potentials are both identical to zero. 

Most importantly, we see that parametrically this expression is similar to the estimate for (the inverse of) Penrose's decoherence time. However, the residual does not scale with the variance of the potentials $\Delta \Phi$, as one might expect from a mean-field approach. Rather it scales with the sum of potentials in each locations. For the interested reader, in the Appendix, I summarise Penrose's calculation and the mean field approach for comparison.

One can rightfully question the details of this argument. The product state might not actually have this form or its time evolution not actually proceed this way. However, this little calculation serves as an estimate for what one generally expects. To maintain a product state, the time evolution in the underlying theory will constantly oscillate into mixed terms ($|\chi_1\rangle | \Phi_2\rangle$ , $|\chi_2\rangle | \Phi_1\rangle$) with an oscillation frequency determined by $m |\Phi_{12}|$. Since we are integrating over the absolute value of the residual, these oscillations will not average out but rather add up over a time $\tau$, and increase as $S \sim \tau m  |\Phi_{12}|$. I will in the following refer to this estimate as the ``Penrose-phase''.

We will now turn to the question of what to do with this estimate.

\subsection{The Collapse}
 \label{ssec:collapse}

Since the integrand of Eq.\ (\ref{eq:res})  is strictly positive, it is clear that the absolute minimum of the action is equal to zero. This is the case iff the time evolution follows the Schr\"odinger equation. The estimate in the previous section then tells us that the Schr\"odinger evolution is a good approximation, so long as the time-integral over the gravitational self energy is small. In this case, the product requirement will not make much of a difference. This is what one expects given that it affects only the gravitational sector. 

However, if we consider a situation in which the state is measured (otherwise, what are we to predict?), then the different branches of the wavefunction will become amplified by the detector. This is what makes a detector a detector: It correlates a quantum state with increasingly more other particles, so that we can eventually read out the result. This means that the total mass that is in a superposition of two places will increase due to the measurement. 

Consider for example that we create a superposition of a photon with a beam splitter and then measure the photon with a photomultiplier. In branch 1, the photon will create a cascade of electrons in photomultiplier 1, whereas photomultiplier 2 remains untouched. For the photon on branch 2, it is instead electrons in photomultiplier 2 that move. Eventually these electrons will create a current and text on a screen that will be read by a human. As that happens, the total mass and energy in a superposition of two locations increases. This is sometimes called a ``Schr\"odinger's cat state" or just a ``cat state".

I want to stress here that the relevant distinction is not that between detector 1 and detector 2. It is rather that between detector 1 that has detected a particle and detector 1 that has not detected a particle, and similar for detector 2. Let me denote the states of the detector on branch 1/2 with $|D_{1/2}^{\rm on} \rangle$ and $|D_{1/2}^{\rm off} \rangle$. Then we start with the (matter) state
\beqn
|D_1^{\rm off} \rangle |D_2^{\rm off} \rangle (\alpha_1 |\chi_1 \rangle + \alpha_2 | \chi_2 \rangle)~,
\eeqn
and end with the entangled state
\beqn
\alpha_1 |D_1^{\rm on} \rangle |D_2^{\rm off} \rangle |\chi_1 \rangle + \alpha_2 |D_1^{\rm off} \rangle |D_2^{\rm on} \rangle | \chi_2 \rangle ~.
\eeqn
The total final state will then be a product state with the corresponding state of the geometry that is now also entangled, albeit only internally and not with the matter. The differences in potential then come from the different locations of particles in the on/off states of the detector. 

Importantly, since this is an entangled state, for identical particles, the total size of the effect does not scale with the square root of the number of particles, it scales linearly. This makes sense intuitively because this is how the gravitational potential scales. Note that the particles in each detector do not need to be a coherent state for this scaling. However, for the estimate in the previous subsection to hold, the particles in either branch need to be dislocated enough so that their wavefunctions no longer overlap. 

Our action principle now must take into account the detector. We then see immediately that any state that will be measured which is not to good approximation in one location (and hence has approximately a classical gravitational field), will generate a very large residual once it hits the detector. This means that if we take into account the amplification by the detector, then the only local time evolutions for which the Schr\"odinger equation is a good approximation is one in which the measurement outcome does not create macroscopic superpositions, in the quantifiable sense that the time-integral over the residual (\ref{eq:res}) remains negligible. 

What will happen instead? For this we have to answer the question of what is the end state of the time evolution of the product state that results in the smallest residual. We have partly answered this question already: The residual will accumulate on any time evolution that is in a superposition of two locations. The more mass (or energy) is in this superposition, the faster it will accumulate. To keep the residual small, a more optimal time evolution is hence one that only briefly (and locally!) violates the Schr\"odinger evolution. This will also make a contribution to the residual, but it will no longer accumulate over time.

This is particularly obvious in our example with the beam splitter. If we measure the photon's path after the beam splitter this would create a macroscopic superposition with a large residual. We would expect that the case in which the photon briefly violates the Schr\"odinger evolution at the beam splitter and locally ``collapses'' into just one arm gives a smaller residual.

What will this residual be? To estimate this we can forget about the specifics of the gravitational sector and its Hamiltonian as it is a general question of going off the Schr\"odinger evolution. Let us just generally consider the case with two branches and a superposition 
\beqn
|\Psi' \rangle = \alpha_1 |1\rangle + \alpha_2 |2 \rangle~.
\eeqn
We will assume that both $|1\rangle$ and $|2\rangle$ fulfil the Schr\"odinger equation and wlog $\rm{Im}(\alpha_1) = 0~,~\alpha_1 \geq 0$.

We then want to know what is the residual for the time-dependent (unitary) rotation into branch 2
\beqn
|\Psi' (t) \rangle = \cos(\theta(t)) |1\rangle + e^{\rm i \varphi } \sin(\theta(t)) |2 \rangle~,
\eeqn
where $\theta(t)$ is a monotonically increasing function of $t$,  $\cos(\theta(t_{\rm s}))\coloneq \alpha_1$, $e^{\rm i \varphi }  \sin(\theta(t_{\rm s}))\coloneq \alpha_2$, $\varphi \in \mathbb{R}$, $\theta_{\rm s} \coloneq \theta (t_{\rm s}) \leq \pi/2$, and $\cos(\theta(t_{\rm e})) = 0$ for some start time $t_{\rm s}$ and final time $t_{\rm e}$. This ansatz is already in the energy gauge, so we get 
\beqn 
\sqrt {|\langle R | R \rangle |} &=& \dot \theta ~, 
\eeqn 
and thus $S = \pi/2 -\theta_{\rm s}$. 

There are two things worth pointing out about this expression. First, it does not depend on how fast $\theta(t)$ evolves, it only depends on the beginning and end value. However, in general such a transition can only proceed locally if it happens within an interaction region. Second, this is the smallest possible value that the action can take on for any such transition because we can always choose the basis vectors so that we have a two-state system, and this rotation is the shortest path between the start and end point. 

Why does the state not collapse once the superposition reaches the detector? Because there is no local interaction which can make this happen. If the particle really were to go several paths but upon measurement was found on only one path, then its mass (and/or energy) would be spread over several places upon arrival at the detector, and it would have to suddenly jump to only one place, violating local energy conservation.\footnote{This problem persists if one slowly collapses the superposition on the way to the detector, as we did in \cite{donadi2020toy}. Such a model can be locally causal in Bell's sense if one is only concerned with the measurement outcome, but generically it still has to propagate energy densities outside the lightcone.}   

What this means is that if the system minimises the action, it will locally collapse into a branch that results in a measurement outcome which is to good accuracy a product state between matter and metric, and pick the branch that had the largest amplitude under the Schr\"odinger evolution. We can therefore identify these product states as the pointer states of the measurement device. When I say they are product states, I do not mean they need to be exact product states. They just need to be close enough to product states so that the Penrose-phase remains $\ll 1$.

The example that we used in subsection \ref{ssec:pen} was that of a position-state measurement. However, the model presented here works for any measurement variable. It only matters that the detector amplifies the state to macroscopic size. Suppose, for example, that we measure an energy eigenstate rather than a position eigenstate. Different energy eigenstates might go the exact same path and hit the detector at the same location. The difference in the gravitational potentials will then come from the necessarily distinct detector response that must eventually reach a macroscopic level, not from having detectors in different locations. 

The only problem is that this, of course, is not what we actually observe. In reality, the outcome of a quantum measurement is not always the pointer state with the largest amplitude. What is missing here is a probabilistic element. We will turn to this next. 

\subsection{Born's Rule}
 \label{ssec:Born}

Superdeterministic theories are hidden variables theories. This means they explain the seeming randomness of quantum mechanics as due to our lack of information about variables $\lambda$ which do not appear in standard quantum physics. To recover a probabilistic theory, therefore, we must now incorporate the hidden variables.

For this, we will assume that the probability of a prepared state to go from an initial state $|\Psi_{\rm s}\rangle\coloneq |\Psi(t_{\rm s})\rangle$ into an end state $|\Psi_{\rm e}\rangle \coloneq |\Psi(t_{\rm e})\rangle$ is determined by random variables $X (|\Psi_{\rm s}\rangle, |\Psi_{\rm e}\rangle,\lambda)$, which are, to a good approximation, independent for each possible end-state. The end state into which the initial state evolves is that with the smallest $X$. 

We define the random variables so that they have rate (inverse mean value)
\beqn
r (|\Psi_{\rm s}\rangle, |\Psi_{\rm e}\rangle) \coloneq e^{-2 A(|\Psi_{\rm s}\rangle, |\Psi_{\rm e}\rangle)}~,
\eeqn
where
\beqn
A(|\Psi_{\rm s}\rangle, |\Psi_{\rm e}\rangle) &:=&  \int_{t_{\rm s}}^{t_{\rm e}} dt \sqrt{\langle R| R\rangle}\sqrt{1- |C(t)|^2}/
|C(t)|~, \nonumber \\
C(t) &\coloneq& |\langle \Psi' | \Psi \rangle |~,
\eeqn
 and so that the probability distribution has maximum entropy, which means it is given by
\beqn
\rho(\lambda|r) &=& 
\left\{
        \begin{array}{ll}
     r e^{- r \lambda } & \mbox{for}\quad \lambda \geq 0 \\
     0 & \mbox{for}\quad \lambda < 0   \end{array} \right. ~. \label{eq:probhv}
\eeqn
This distribution depends on the end state $|\Psi_{\rm e} \rangle$ via $r$ and hence on the measurement settings and thereby violates measurement independence. 

To understand this expression, we note that for the rotation in section \ref{ssec:Teleology}, we have
 $C(t) =  \cos(\theta(t) - \theta_{\rm s})~$, so that
\beqn 
A &=& \int_{\theta_s}^{\pi/2}  d \theta  ~ \frac{\sin(\theta  - \theta_{\rm s})}{ \cos(\theta - \theta_s)} \nonumber \\&=& - \ln(\sin (\theta_{\rm s})) = - \ln |\alpha_2| ~,
\eeqn
and hence $r=|\alpha_2|^2$.

In the case of the enforced product state of section \ref{ssec:Off} we have instead
\beqn
C(t) = | \alpha^2_1 {\alpha'_1}^*   + \alpha^2_2 {\alpha'_2}^* | ~. \label{eq:cp}
\eeqn
Its absolute value generically gives a correction of order one. That is, by order of magnitude we get
\beqn
A \sim \tau m |\Phi_{12}|~,
\eeqn 
where $\tau$ is the time that the superposition remains in existence.

 To move on, we define $\Delta t = t_{\rm e} - t_{\rm s}$ and $|\Psi'_{\rm e} \rangle := e^{-{\rm i} H \Delta t} |\Psi_{\rm s} \rangle$ as the final state under the Schr\"odinger evolution.  We have previously seen that if the final state is to good accuracy a product state of metric and matter it can serve as a measurement eigenstate. We will denote these states as $|\Psi_I \rangle$, $I \in \{ 1,2\dots D \}$, and $\alpha_I \coloneq \langle \Psi_I | \Psi'_{\rm e} \rangle$.

Let us then look at the case with two final states $|\Psi_{\rm e1} \rangle$ and $|\Psi
_{\rm e2} \rangle$. We want to know what is the probability that we get the final state $|\Psi_{\rm e1} \rangle$. By assumption this will be the case whenever
\beqn
X(|\Psi_{\rm s}\rangle,|\Psi_{\rm e1} \rangle, \lambda_1) <  X(|\Psi_{\rm i}\rangle,|\Psi_{\rm e2}\rangle, \lambda_2)~,
\eeqn
that is, the values have to fulfill the condition $\lambda_{1} < \lambda_{2}$. 

We can then calculate the probability of this happening by integrating over the probability distributions
\beqn &&
P(X(|\Psi_{\rm s}\rangle,|\Psi_{\rm e1}\rangle, \lambda_1) <   X(|\Psi_{\rm s}\rangle,|\Psi_{\rm e2}\rangle, \lambda_2)) = \nonumber \\
&& {r_1 r_2} \int_0^\infty d \lambda_{1}   \int_{\lambda_{\rm 1}}^\infty d \lambda_2 ~e^{- r_1 \lambda_1 -  r_2 \lambda_2}~,
\eeqn
where $r_1$ and $r_2$ are the rates associated with end state 1 and 2, respectively.
 The result of this integral is
\beqn
\frac{r_1}{r_1 + r_2} = \frac{e^{-2A(|\Psi_{\rm s} \rangle, |\Psi_{\rm e1} \rangle)}}{e^{-2 A(|\Psi_{\rm s} \rangle, |\Psi_{\rm e1} \rangle)}  + e^{-2A(|\Psi_{\rm s} \rangle, |\Psi_{\rm e2} \rangle)}}~. \label{p3}
\eeqn
We see that if the end state $|\Psi_{e1} \rangle$ is that of the usual Schr\"odinger evolution but not a product state, then the probability is exponentially suppressed. We hence never observe these outcomes. The best possible path is one with a sudden branch rotation. In this case $P = |\alpha_1|^2$ for the state to go to end state $|\Psi_1 \rangle$, which is what Born's rule requires. 

Do the probabilities come out correctly for arbitrary numbers of $|\Psi_I \rangle, I \in \{ 1,2,\dots D \}$? Yes, they do. One can confirm this the hard way by actually calculating the integral for $D$ probability distributions. But we can make it simpler by denoting $P_I:=P(|\Psi_{\rm e}\rangle = |\Psi_I\rangle)$.
It follows from the above that
\beqn
\frac{|\alpha_I|^2}{|\alpha_I|^2 + |\alpha_J|^2} = \frac{P_I}{P_I + P_J}\quad \forall \quad I \neq J~.
\eeqn
Since this equation is trivially also true for $I=J$, we get
\beqn
\frac{|\alpha_I|^2}{|\alpha_J|^2} = \frac{P_I}{ P_J}\quad \forall \quad I, J~.
\eeqn
And because both the $P_I$ and the $|\alpha_I|^2$ sum up to 1, we have
\beqn
1 = \sum_{I=1}^D P_I = \frac{P_J}{|\alpha_J|^2} \sum_{I=1}^D |\alpha_I|^2 = \frac{P_J}{|\alpha_J|^2} ~.
\eeqn
So, $P_J = |\alpha_J|^2 ~ \forall ~ J$. 

We also see that since the integral in $A$ is additive and in the exponent, the probabilities of consecutive local rotations will factorise, as they should. 

A few comments. One might raise the point here that the probability distribution of the hidden variables (\ref{eq:probhv})  appears like a bunny out of a magic hat rather than following from anything in particular. This is entirely true, but absent an underlying theory really the best I can do is to provide an example for the probability distribution that reproduces quantum mechanics\footnote{And that, as we may note, does not require finetuning, thus providing another counterexample to the claim that superdeterministic theories are finetuned.}. For what I am concerned, it doesn't pop any more out of the hat than Born's rule pops in the standard formulation of quantum mechanics. Of course one hopes that eventually one would be able to derive this (or some other) expression from some deeper theory, but we will leave this to future work\dots 

A second question one might have is why I introduced both $S$ and $A$ and not just used $A$ for both purposes. The reason is not only that this doesn't make physical sense, it doesn't work. $S$ is the functional whose variation determines the possible paths that the system can take. $A$ is a statistical measure that counts paths in the presumed to be underlying theory. They should not be the same. 
Worse, if we set them to be the same, we would run into the problem that $\delta A=A=0$ on any path of the form $e^{i {\varphi(t)}} |\Psi' \rangle$, $\varphi(t) \in \mathbb{R}$. That is, we would lose the usual Schr\"odinger evolution as a unique solution in the case when gravity can be neglected.

Another question one might have here is how this can possibly reproduce Born's rule if there are other time evolutions from the initial to the final state that we could realise with local interactions even disregarding gravity. 
To be concrete, consider a Mach-Zehnder interferometer. Under the usual Schr\"odinger evolution, the state goes in a superposition over both paths, and then recombines. But we could alternatively use a local rotation at the first beam splitter to either one of the paths, and then another local rotation on the second beam splitter into the wrong port. Why does this not happen?

The reason this cannot happen is that such a path would not be a stationary solution of $S$. As long as the outcome of the usual Schr\"odinger evolution is a detector eigenstate, the product state constraint does nothing and the usual solution of the Schr\"odinger equation is the optimal path. It is only when the forward evolution of the Schr\"odinger solution is not a product state between matter and geometry that the product states become the stationary solutions.  

In case you noticed that (\ref{eq:cp}) could be equal to zero for $\alpha_{1/2} = \alpha'_{1/2}$, $|\alpha_1| = |\alpha_2| = \sqrt{1/2} $, and a suitably chosen phase, do not worry. This path is not stationary since any nudge towards smaller $|\alpha_1|$ or $|\alpha_2|$ will decrease the residual, and there always must be a local path for this, since otherwise the state that we actually measure could not come about locally. That is, it was somewhat unnecessary to include this case here. However, it will make it easier to interpret the mathematics, which we turn to next. 

\subsection{Interpretation}
\label{ssec:int}

Before we get to the experimental tests I would like to offer a hopefully helpful interpretation for what is going on. In the framework developed here, a massive object that is in a superposition of two different locations is somewhat like a virtual particle-antiparticle pair in a Feynman diagram. It can exist temporarily but will not appear in outgoing states, just that here the ``outgoing'' states are detector eigenstates that must be, to good precision, product states of matter and metric. The required precision is given by the accumulated Penrose phase. 

However, the temporal and spatial extension of these intermediate superpositions much exceeds those of virtual particles because quantum gravitational effects are so small. They can exist over a duration given by approximately $\tau \sim 1/(m |\Phi_{12})|$ which, for elementary particles, is enormously large (we will get to estimates in the next section). 

The teleological element of this construction is that the question of whether the state stays in a superposition or collapses depends on whether it {\emph{will}} be measured in the future, i.e.\ whether it will go on to interact with a detector or not. 

Personally I do not read much into this mathematical property because I do not think of this model as fundamental. There is also no good reason for why virtual particles can't become real other than that we know they are just a way to keep track of integrals, and real particles must be on-shell by construction. Asking how virtual particles ``know'' that they need to disappear again is just a meaningless question. I think that the question of how a particle in a superposition of two locations ``knows'' it must recombine is equally meaningless. But the reader who feels uncomfortable with the future-dependence might want to imagine that indeed all possible time evolutions happen, each in its own universe, it's just that the probability that we find ourselves in a universe with a cat-state is vanishingly small, as we saw in the previous subsection.

The comparison to virtual particles also helps to understand why, in this approach, we do not need to integrate the residual all the way to infinity. This is because for practical purposes we can chop the time evolution apart into disconnected diagrams at any sufficiently localised and near classical (think ``real'') final state, that is the end of a measurement process.

 \subsection{Weak Measurements}
 \label{ssec:weak}

 Once we have the probabilistic formulation, we can deal with weak measurements. One of the standard definitions for weak measurements is loosely speaking a detector that only sometimes makes a detection. Concretely, let us consider a weak measurement that answers the question ``Is the system in state $|q \rangle$?'' The detector can then be described by two operators $\hat M_{+}$ and $\hat M_{-}$, the former describing a positive, the latter a negative (no) detection. These operators can be defined as
 \beqn
 \hat M_+ &\coloneq& \sqrt{p} ~|q\rangle \langle q|~, \\
 \hat M_- &\coloneq& \openone - \left(1- \sqrt{1-p} \right) ~ |q\rangle \langle q|~,
 \eeqn 
 with some detection probability $p \ll 1$.  That is, we reformulate the weak measurement as a collapse that happens only with a small probability \cite{jacobs2014quantum}.

 There is a simple way to interpret such a device in the model presented here. A weak measurement detector is one that only sometimes maintains the residual $||R||$ and sometimes washes it out. This can happen if the imprint of the state $|q\rangle$ was not large enough for the detector to reliably amplify it to a macroscopic level. An example could be that frequent interactions with atoms in a gas can make states less, rather than more, distinguishable, essentially by adding noise. Whether a system actually acts as a detector, therefore, depends not only on its number of constituents, but on how they interact and just how quickly that increases $||R||$. A weak measurement device sits at the threshold between detection and no detection.

 A second definition of weak measurement which is frequently used \cite{aharonov1988result} is that the detector itself is described by a state $|M\rangle$ that changes only mildly to a state $|M'\rangle$ with large overlap, i.e.\ $|\langle M|M' \rangle| \sim 1 $. In this case, the `weakness' of the measurement lies in the difficulty of reading out the difference between the two states, which only sometimes works. 
 
 The distinction between the two definitions of a weak measurement is therefore about what one calls the ``measurement'' and not the actual mechanism. I think that the imprint in $|M' \rangle$ would better be referred to as a pre-measurement and the term measurement be reserved for a successful amplification. In any case, both definitions mean that the record of the observable is only sometimes amplified to macroscopic scales, which can be described in the context of this model as a system that only sometimes accumulates a large Penrose phase.

 One must keep in mind though that the question of whether a device acts as a detector does not depend on whether $||R||$ grows, but whether $\int dt ||R||$ reaches $ \sim 1$. That is, the detection threshold is a residual that grows faster than $1/t$ for a sufficient duration.

\subsection{Free Particles}
\label{ssec:free}

In the previous subsections we assumed that the wavefunctions on each branch are gravitationally coherent states. By this I mean that the matter $\otimes$ gravity sector is coherent. The matter and gravity sectors do not have to be each coherent internally. Gravitationally coherent states will, by construction, not build up a residual.

Canonically coherent states in quantum mechanics are defined as eigenstates of the annihilation operators. They are often interpreted as the closest approximation to a classical state. Their main feature is, as the name suggests, that they have only one overall phase.  
However, to avoid an accumulation of the Penrose-phase it is sufficient that a state remains coherent, it need not necessarily be a canonically coherent state: any state that maintains only one overall phase under the canonical evolution will do. 

If one strictly defines a freely propagating particles as one that asymptotically goes to infinity, then any arbitrarily small residual would build up and eventually exceed the residual cost of going off the Schr\"odinger evolution at the previous interaction. Therefore, in the model presented here, asymptotic particle states in  Feynman diagrams must be gravitationally coherent. Then again, we never observe asymptotic states, because to observe them they must go into a detector and not to infinity.

\section{Tests}

Before we turn to estimates for the prospects of testing the model proposed here, I want to point out some relevant differences to the Penrose-Di\'{o}si (hereafter PD) model. 

In the PD model, the deviation from the standard Schr\"odinger equation comes from a noise-kernel that scales with the gravitational self-energy of the mass-density. The strength of this effect first grows for small separation, as long as the wave-packets overlap. But once the wave-packets are to good approximation orthogonal, the strength of the deviation drops with the inverse of the distance between them. (If the branches are not each internally coherent, a residue remains for each branch.) 

In the model proposed here, in contrast, the contribution to the Penrose phase grows with the dislocation of the wave-packets, but once the wave-packets are orthogonal it becomes constant and to good approximation independent of the distance\footnote{Strictly speaking, there is a contribution to the phase from the Newtonian potential at one branch evaluated at the other. However, I set that to approximately zero exactly because it will drop with separation.}. This is basically because the residual that I am using measures distance in Hilbert-space and not in spacetime. 

The other difference between the models is what I already mentioned in \ref{ssec:pen}: The effect in this new model scales with the sum of the gravitational potentials, whereas in the PD-model it scales with the variance of the gravitational self-energy. In both cases though one has to avoid integrating over point-like sources because that would bring in a divergence from the Newtonian potential. I will therefore here, as usual, assume that the mass-density is smeared out over a radius that roughly scales like the width of the wave-packet.

\subsection{Estimates}
\label{ssec:testes}

The great benefit of this model is that since the collapse is caused by a known entity---gravity---the process has no free parameters. Let us therefore make some order-of-magnitude estimates for when the effects of the model proposed here should become relevant. 

In the best case (near a maximally localised particle), we can estimate $\Phi_{12} \sim (m/m_{\rm p})^2$, where $m_{\rm p}$ is the Planck mass.  This gives us a Penrose phase $\sim \tau m^3/m_{\rm p}^2$. Our task is now to find cases where the collapse induced by this phase is not masked by decoherence. 

For elementary particles, the Penrose phase is ridiculously small. For an electron, for example, its effect will become noticeable at a time of approximately
\beqn
\tau_{\rm e}  \sim \frac{m_{\rm p}^2}{m^3} \sim 7 \times 10^{23} ~ \rm{sec}~. \label{T1}
\eeqn
Even for a heavy nucleus with, say, 100 nucleons, we have $m \sim$ 100 GeV and $\tau \sim 10^8$~seconds. These effects will be masked by environmental decoherence easily. 

What about quantum computers? At first, this sounds like a promising idea because quantum computers are designed to produce massive amounts of entanglement while keeping decoherence at bay. However, in a quantum computer the masses that move are tiny.

Consider for example a superconducting circuit that will move something like $N_{\rm e}=10^6$ electrons. 
To get a collapse time of $\sim 1$~second, we would thus optimistically need $\sim 10^{17}$ qubits, if we were to create a fully entangled state. To take the other extreme, if we just have a product of qubits, each of which is in a superposition (of electron locations), then we have to use the $\sqrt{n}$ scaling from  subsection \ref{ssec:mult} and we need $\sim 10^{35}$ qubits. In a realistic setting the number would be somewhere in between but probably closer to the latter estimate. In any case, clearly, this is not going to happen any time soon. The situation is even worse for other types of qubits, such as ion traps, neutral atoms, or photon states, because they dislocate even smaller masses. This makes my estimate considerably less optimistic than those put forward in \cite{t2016cellular,Palmer:2025guq}. 

What we need is instead a lot of mass that moves coherently, so that we can witness its collapse. This is the bad news. The good news is that this mass does not need to move by a lot, it just needs to move enough so that we can consider the wavefunctions as sufficiently displaced for our estimate to hold. This is not much. We just need to displace e.g.\ atomic nuclei by more than the typical diameter of the nucleus, that is, some femtometres. 

\subsection{Existing Proposals}
\label{ssec:testex}

Experimental setups which are in the parameter range of testing the model proposed here (maybe not so surprisingly) are attempts to probe quantum gravity by bringing small objects (typically made of silicon) into a superposition of two coherent oscillation states \cite{belenchia2018quantum,bild2023schrodinger,bose2025massive}. To get a collapse time of $\sim 1$ second, we would need to displace a total mass of about $0.2$ nanogram (or, equivalently, a mass of about 1 ng with a coherent fraction of $0.2$). The model proposed here predicts that superpositions which exceed this bound from the Penrose phase do not exist, or they cannot stay coherent, respectively. The decoherence time of these objects is currently in the ms range \cite{halg2022strong}, so either the masses of the oscillators need to further increase or the coherence be improved, but we are not so far away from being able to test this model. I consider this to be the currently most promising experimental avenue. 

A completely different way to test this model would be to see whether matter and gravity actually can be entangled. There are some experiments gearing up to look for this, see e.g.\ \cite{carlesso2019testing}. However, most of the experiments currently underway that use ``entanglement witnesses'' \cite{bose2017spin,marletto2017gravitationally} to probe the quantisation of gravity actually measure the entanglement indirectly through the matter sector, so they are not sensitive to the product state constraint.  

The product state requirement will not affect real graviton emission provided one treats the graviton also as a particle\footnote{The assumption that particles and geometry are ultimately the same does not per se require the existence of gravitons. However, I find it hard to see how one could have geometry with quantum features and not also have propagating modes, i.e.\ gravitons.}. It will set constraints on the possible matter states created in scattering processes. However, since we have zero evidence that gravitons exist, and measuring them is far outside experimental reach for the foreseeable future, it is rather moot to discuss this point. 

Another possibility to test this model would be to investigate closer just which type of in-medium interactions result in a sufficient accumulation of $||R||$ to induce a collapse, and whether there are cases when that would happen before decoherence makes the effect undetectable. However, for all in-medium interactions that I have looked at so far, the accumulated residual and the inverse decoherence time are pretty much the same, so this does not seem to be promising.

It is worth noting that the model proposed here {\emph{cannot}} be tested by looking for dispersive effects like those induced by the noise-kernel of the Penrose-Di\'{o}si model \cite{donadi2021underground}. 

\section{Discussion}

I now want to spend a few words on why I think it makes sense to treat gravity differently from the other interactions. 

In canonically or perturbatively quantised gravity, the entanglement between matter and its gravitational field can be best understood by what is called a `dressing' in the context of quantum field theory. This is the formal acknowledgement that a particle which carries a charge---say, an electron that carries electric charge---never occurs in nature without the field created by that charge. A bare electron appears only in the mathematics. Real electrons always come with soft (low energy) photon clouds: This is the dressing. 

If one has an electron that is in a superposition of two branches, then each branch has its own dressing. They are independent from each other and can have a relative phase. Consequently, the electron is entangled with the soft photon dressing. The entanglement in quantum gravity comes about in the same way: it describes a dressing by soft gravitons that can then be interpreted as the field caused by the particle, or the geometric deformation respectively. 

In the approach that I started from, however, soft photons are very different from soft gravitons. The photons are particles and independent of the electrons. The soft gravitons, however, are not because, by assumption, the particle and its geometry are ultimately the same. They are the same, that is, up to freely propagating modes that are hard gravitons. 

The reason this makes sense at least to me is that if we treat the graviton dressing like the photon dressing, then the mathematical description of a particle in a superposition of multiple locations carries no information about it in fact being {\emph{one}} particle. There is formally no way we could know that, say, half an electron with its own dressing is actually not itself a fundamental particle other than by normalisation. But the normalisation just tells us that we cannot measure only one branch in isolation, which is exactly the fact that needs explanation. But half an electron does not exist any more than an electron without an electric field. To put this differently: Two branches of one particle are not physically independent because they cannot exist separately. 

The product state requirement can thus be seen as a way to make sure that it is always {\emph {entire}} particles that create a geometry, and this in turn explains why we can only measure entire particles. In other words, the Hamiltonian constraint should act per particle sector (of Fock space), and not point-wise. This is why, intuitively, the state of a particle in two branches has mixed terms that essentially describe part of a particle in one place with a gravitational field sourced from another place. This happens exactly because the rest of the particle must be somewhere. 

I know that I did not, in fact, put forward the mathematics for these statements. That is because it would just add assumptions that are unnecessary to arrive at the phenomenological consequences which were the focus of this paper. However, I wanted to provide this explanation as motivation. 

It is also worth mentioning that the estimate presented here does not rest on the product state assumption. By order of magnitude one would expect most deviations from canonically quantised gravity to give a similar result.

I interpret this model as a way to reconcile the present formalism of quantum mechanics with a possibly underlying theory. The teleology of this model is likely a mathematical artifact that originates in the way that we have developed quantum mechanics. This is because the teleology of the model presented here ultimately comes from the way that we describe the initial (prepared) state of the system. But how do we even know what this state is? We have simply inferred these states (and their Hamiltonian operators) from the results of countless experiments. We have no evidence that the initial state in standard quantum mechanics is an ontological state, rather, it is a reconstruction for the purpose of fitting our measurement results. 

Is it any surprise, then, that if we ask for the local time evolution of such a reconstructed state, we get a model in which the choice of measurement variable must fit with what happened before the measurement? I think that this odd feature will make sense once we understand the underlying physics, just like the electron's magnetic moment made sense once we understood what a spin $1/2$ particle is.

Finally, I want to stress  that nowhere have I assumed that the geometry is classical. Indeed, in general it is not classical: The geometry can have superpositions and it can also be internally entangled. It is just that states of the geometry with significant quantum features will not survive for long because of the residual build-up.  

\section{Summary}

I have shown here how the assumption that matter and geometry have the same fundamental origin requires the time evolution of a quantum state to differ from the Schr\"odinger equation. This has the consequence that the ideal time evolutions which minimise the action are those with end states that are to good approximation classical. We can then identify these end states with the eigenstates of the measurement device. This new model therefore explains why quantum states seem to `collapse' into eigenstates of the measurement observable, and how this can happen while preserving locality. 

Since the collapse process is governed by quantum gravitational contributions whose strength is known, the resulting model is parameter free. Collapse happens in this model whenever the accumulated phase difference between dislocated branches, $\tau m | \Phi_{12}|$, exceeds $\sim 1$. The model's phenomenology---notably the collapse itself---can be tested in roughly the same parameter range as other tests of the weak field limit of quantum gravity. 

 \bigskip
 
\textit{Acknowledgements} 

\bigskip

I acknowledge help from ChatGPT 5 for literature research as well as checking this manuscript. I swear I actually wrote it myself. 

\section*{Appendix}

I here want to briefly summarise Penrose's argument \cite{penrose1996gravity} and the mean field approach to semi-classical gravity. As in the body of this paper, we consider a particle in a superposition of two locations $\vec x_1$ and $\vec x_2$. Each has an energy density, $\rho_1$ and $\rho_2$, centred around its position,  but the particles are so far apart that their overlap is negligible. We will assume that each lump of the superposition on its own is to good accuracy classical and denote their Newtonian potentials as $\Phi_1$ and $\Phi_2$, respectively, where $\nabla^2 \Phi_{1/2} = 4 \pi G \rho_{1/2}$.

In approaches like perturbatively or canonically quantized gravity, the full state will be entangled with the metric, schematically $ \alpha_1 |\chi_1 \rangle |g_1\rangle + \alpha_2 |\chi_2 \rangle |g_2 \rangle$, where $|\chi_{1/2} \rangle$ are the parts of the matter wave-function in either location, and $|g_{1/2} \rangle$ the metrics. 

Now remember that in General Relativity the Newtonian potential appears as a correction to the time-time component of the metric tensor $g_{tt} \approx - ( 1 + 2 \Phi)$. Also remember that the lapse $N$, which can be said to quantify the passage of time, is basically $N = \sqrt{-g_{00}} \approx 1 + \Phi$. We can then estimate the mismatch between the two time coordinates as $\delta N \sim (\Phi_1 - \Phi_2) \coloneq \delta \Phi$. This causes a discrepancy in the local acceleration of order $\delta a \sim - \nabla \delta \Phi$. 

The important step is now that Penrose interprets this discrepancy as a measure for the \emph{uncertainty} in the time-evolution, and not just a difference. He argues that a natural measure of this uncertainty is the contribution to the gravitational self-energy which is
\beqn
 E^{\rm (Pen)}_{\rm G} \coloneq \frac{1}{4 \pi G} \int d^3 x~ (\nabla \delta \Phi )^2 ~.
\eeqn
Integrating by parts and discarding the boundary term gives
\beqn
 E^{\rm (Pen)}_{\rm G} = - \frac{1}{4 \pi G} \int d^3 x~ \delta \Phi (\nabla^2 \delta \Phi ) ~.
\eeqn
We can then insert the non-relativistic propagator
\beqn
\Phi(\vec x) = - G\int d^3 y~ \frac{\rho(\vec y)}{|\vec x - \vec y|}
\eeqn
and get
\beqn
 E^{\rm (Pen)}_{\rm G} = \frac{G}{2} \int d^3 x \int d^3 y  \frac{ \delta \rho (\vec x) \delta \rho(\vec y)}{|\vec x-\vec y|} ~, \label{eq:deltaE}
\eeqn
where $\delta \rho \coloneq \rho_1 - \rho_2$. 

Since Penrose interprets the spread in lapses as an uncertainty, he says that this self-energy mismatch will cause different branches to lose phase-correlations at a timescale of $\tau \sim 1/ E^{\rm (Pen)}_{\rm G}$.  He then conjectures that ``The {\textit{basic stationary states}}
into which a general superposition would decay by state reduction are to be stationary
solutions of the \emph{Schr\"odinger-Newton equation} (SN-equation),'' (\cite{penrose1998quantum}, emph orig.) which is similar to the behaviour we found in the previous subsections.

However, such a decoherence just does not happen in either canonically or perturbatively quantised gravity. Part of my reason for writing this paper was to close this gap.

One finds a similar result if one considers the Schr\"odinger-Newton equation \cite{ruffini1969systems,van2011schrodinger}
\beqn
{\rm i} \partial_t \psi(\vec x, t) &=& \left( - \frac{1}{2m} \nabla^2 + V + m ~\Phi_N \right) \psi(\vec x, t) 
\eeqn
with
\beqn
\nabla^2 \Phi_{\rm N} &=& 4 \pi Gm~ |\psi(\vec x, t)|^2   ~,
\eeqn
as a mean-field approximation. How good this approximation is can be quantified by calculating the variance of the density, that is, the strength of the quantum fluctuations around the mean value, where the mean value gives rise to the classical metric. For this, one replaces $\rho$ with an operator $\hat \rho$ that is acting on the matter state, and defines $\delta \hat \rho \coloneq \hat \rho - \langle \hat \rho \rangle$ (note: this operator is {\bf not} the same as the function $\delta \rho$ above), then $\langle \delta \hat \rho ~\delta \hat \rho \rangle = \langle \hat \rho^2 \rangle - \langle \hat \rho \rangle^2$ is the variance and 
\beqn
E^{\rm (var)}_{\rm G} = \frac{G}{2} \int d^3 x \int d^3 y~ \frac{\langle \delta \hat \rho (\vec x) ~ \delta \hat \rho(\vec y) \rangle}{|\vec x - \vec y|} ~ \label{eq:deltaH}
\eeqn
measures the deviations from the mean field approximation.

\bibliographystyle{naturemag}
\bibliography{ref.bib}

\end{document}